\documentclass[a4paper,11pt]{article}
\usepackage{pos}
\usepackage{soul}
\usepackage{orcidlink}

\usepackage{amsthm}
\usepackage{slashed}
\usepackage[english]{babel}
\usepackage[utf8]{inputenc}
\usepackage{graphicx}
\usepackage{color}
\usepackage{amsfonts,amsthm}
\usepackage{bm,bbm}
\usepackage{xcolor}
\usepackage{lipsum}
\usepackage{comment}

\newcommand{\cA}{\ensuremath{\mathcal A} }

\newcommand{\cgrav}{\ensuremath{c_{\text{grav}}} }
\newcommand{\cD}{\ensuremath{\mathcal D} }

\newcommand{\cF}{\ensuremath{\mathcal F} }

\newcommand{\cN}{\ensuremath{\mathcal N} }

\newcommand{\vn}{\mathbf{n}}

\newcommand{\cO}{\ensuremath{\mathcal O} }
\newcommand{\cP}{\ensuremath{\mathcal P} }
\newcommand{\cQ}{\ensuremath{\mathcal Q} }

\newcommand{\cU}{\ensuremath{\mathcal U} }

\newcommand{\al}{\ensuremath{\alpha} }
\newcommand{\ga}{\ensuremath{\gamma} }

\newcommand{\la}{\ensuremath{\lambda} }

\newcommand{\refcite}[1]{Ref.~\cite{#1}}

\newcommand{\eq}[1]{Eq.~\ref{#1}}

\title{Lattice Studies of Two-Dimensional Maximally Supersymmetric Yang--Mills Theory for Tests of Gauge--Gravity Duality}
\ShortTitle{Lattice Studies of 2D MSYM Theory for Tests of Gauge--Gravity Duality}

\author*[a]{Bana Singh Sangtan~\orcidlink{0009-0008-5841-873X}}
\affiliation[a]{Department of Physical Sciences, Indian Institute of Science Education and Research - Mohali, Knowledge City, Sector 81, SAS Nagar, Punjab 140306, India}
\emailAdd{bsangtan@gmail.com}

\author[b]{Anosh Joseph~\orcidlink{0000-0003-4288-8207}}
\affiliation[b]{National Institute for Theoretical and Computational Sciences, \\ School of Physics, and Mandelstam Institute for Theoretical Physics,\\ University of the Witwatersrand, Johannesburg, Wits 2050, South Africa}
\emailAdd{anosh.joseph@wits.ac.za}

\author[c]{David Schaich~\orcidlink{0000-0002-9826-2951}}
\affiliation[c]{Department of Mathematical Sciences, University of Liverpool, Liverpool L69 7ZL, United Kingdom}
\emailAdd{david.schaich@liverpool.ac.uk}

\FullConference{The 42nd International Symposium on Lattice Field Theory (LATTICE2025)\\
2-8 November 2025\\
Tata Institute of Fundamental Research, Mumbai, India\\}

\abstract{
We present our ongoing work on two-dimensional maximally supersymmetric Yang--Mills (2D MSYM) theory using lattice techniques. 
The continuum theory is obtained from the dimensional reduction of four-dimensional ${\mathcal N} = 4$ supersymmetric Yang--Mills theory. 
We construct both the continuum and lattice versions of the 2D MSYM theory. 
The lattice action preserves a subset of supersymmetries. 
We extend existing lattice software with new routines to accommodate the additional terms in the lower-dimensional theory. 
This lattice construction enables us to perform Rational Hybrid Monte Carlo simulations of 2D MSYM and facilitates the exploration of its continuum limit. 
Our work contributes to the numerical study of maximally supersymmetric gauge theories and supports the ongoing efforts to test gauge--gravity duality and investigate related non-perturbative phenomena.
}

\begin{document}
\maketitle

\section{Introduction}

Maximally supersymmetric Yang--Mills (MSYM) theory in $(p+1)$ dimensions plays a central role in gauge/gravity duality. 
In the large-$N$ limit, $(p+1)$-dimensional $\mathrm{SU}(N)$ MSYM is conjectured to be dual to Type~IIA (even $p$) or Type~IIB (odd $p$) string theory with $N$ coincident D$p$-branes in an appropriate decoupling limit~\cite{Itzhaki:1998dd}.
The best-known example is the $p = 3$ case, four-dimensional $\cN = 4$ SYM, which yields the original AdS/CFT correspondence~\cite{Maldacena:1997re}. 
For $p \neq 3$, the theories are non-conformal but still admit well-defined holographic descriptions. 

In this work, we focus on the $p = 1$ case: two-dimensional (2D) MSYM at finite temperature. 
We present its lattice formulation for a non-perturbative study.
The spatial direction is compactified with periodic fermion boundary conditions, while the Euclidean temporal circle implements finite temperature through anti-periodic fermion boundary conditions. 

At low temperature and large $N$, the dual description is given by supergravity solutions corresponding to charged black objects.
Two phases arise: homogeneous black strings wrapping the spatial circle and localized black holes~\cite{Aharony:2004ig, Aharony:2005ew}.
These are separated by a first-order Gregory--Laflamme transition~\cite{Gregory:1993vy}.
Holography predicts that this transition is dual to a spatial deconfinement transition in the gauge theory, with the spatial Wilson line serving as the order parameter. 

Because analytic control is limited at strong coupling, lattice field theory provides a natural non-perturbative framework. 
Early lattice studies of 2D MSYM observed evidence for the expected transition at small $N \leq 4$~\cite{Catterall:2010fx}, and later work extended this to $N \le 16$ and larger lattice volumes~\cite{Catterall:2017lub}.
In particular, Ref.~\cite{Catterall:2017lub} employed software designed for four-dimensional $\cN = 4$ SYM, and the ongoing work we report here aims to improve upon this by directly implementing the two-dimensional theory.
This should make it possible to explore larger $N$ and finer lattices, improving control over the large-$N$ and continuum limits.

The thermal theory is formulated in Euclidean signature on a torus, where the temporal cycle has circumference $\beta$ and anti-periodic fermion boundary conditions. 
Importantly, the lattice construction employed here naturally yields skewed tori rather than purely rectangular geometries. (The skewed torus reduces to rectangular one for certain aspect ratios.) 
Modern twisted lattice formulations of $d$-dimensional theories with at least $2^d$ supersymmetries preserve at least one exact supersymmetry at non-zero lattice spacing, and require non-hypercubic lattices that are naturally adapted to such skewed toroidal backgrounds.
See Ref.~\cite{Schaich:2022xgy} for a recent review.

\section{Review of thermal large-$N$ 2D MSYM on a circle} 
\label{sec:gravity} 

In this section, we give a summary of the theoretical expectations for large-$N$ 2D MSYM compactified on a spatial circle of circumference $L$ at temperature $T = 1/\beta$. 
These results combine field-theoretic arguments with holographic input from the dual gravity description. 

The thermal theory is formulated in Euclidean signature with $\tau \sim \tau + \beta$, so that it lives on a flat rectangular torus with side lengths $\beta$ and $L$. 
Fermions obey anti-periodic boundary conditions along the temporal circle and periodic boundary conditions along the spatial circle. 

The Euclidean action is
\begin{equation}
S = S_{\text{Bos}} + S_{\text{Ferm}},
\end{equation}
\begin{equation}
S_{\text{Bos}} = \frac{N}{\lambda} \int d\tau\, dx\ {\rm Tr} \left(\frac{1}{4} F_{\mu\nu} F^{\mu\nu} + \frac{1}{2} (D_{\mu} X^I)^2 - \frac{1}{4} [X^I, X^J]^2 \right),
\end{equation}
\begin{equation}
S_{\text{Ferm}} = \frac{N}{4\lambda} \int d\tau \, dx \, {\rm Tr} \left(\Psi \left(\slashed{D} - [\Gamma^I X^I, \cdot] \right) \Psi \right),
\end{equation}
where $X^I$ ($I = 2, \ldots, 9$) are eight adjoint scalar fields arising from dimensional reduction of ten-dimensional $\cN = 1$ SYM, and $\Psi$ is the corresponding adjoint fermion.

The 't~Hooft coupling $\la = N g_{\text{YM}}^2$ is dimensionful. 
We define dimensionless parameters $r_L = L \sqrt{\lambda}$, $r_\beta = \beta \sqrt{\lambda}$ and $t = 1 / r_\beta$, and consider the large-$N$ limit with $r_\beta$ and $r_L$ fixed. Key observables are thermodynamic quantities derived from the bosonic action density and the normalized Wilson lines
\begin{equation}
\label{eq:Pdefn}
P_{\beta,L} = \frac{1}{N} \left\langle \left| {\rm Tr} \, \cP e^{i\oint_{\beta,L} A} \right| \right\rangle,
\end{equation}
wrapping the thermal and spatial cycles.
At large $N$, these serve as order parameters for $\mathbb{Z}_N$ center symmetry breaking: $P_\beta = 0$ ($\neq 0$) indicates thermal confinement (deconfinement), and similarly $P_L = 0$ ($\neq 0$) signals spatial confinement (deconfinement).

In the following, we review the expected phase structure in the $(r_\beta, r_L)$ plane based on field-theoretic limits and holographic predictions. 

\subsection{Dual gravity description at low temperature} 
\label{sec:dualGrav} 

In the large-$N$ and low-temperature regime, $r_\beta \gg 1$, holography predicts a dual description in terms of D1-charged supergravity solutions in Type~IIB string theory~\cite{Itzhaki:1998dd}.  
The supergravity approximation requires $N \gg 1$ and $r_\beta \gg 1$ in order to suppress string quantum and $\alpha'$ corrections. Also, $r_L \ll r_\beta$ is needed to avoid winding effects along the spatial circle. 

In this regime, the dominant configuration is a homogeneous black string wrapping the spatial circle. 
In the gauge theory, this \emph{D1 phase} predicts the free energy density
\begin{equation}
\label{eq:D1phase}
\frac{f_{\text{homog}}}{N^2 \lambda} \simeq - 3.455 \, t^3.
\end{equation} 

The above solution becomes dynamically unstable and a Gregory--Laflamme (GL) instability sets in for $r_L^2 \le c_{\text{GL}} \, r_\beta$, where $c_{\text{GL}} \simeq 2.24$.
Below this threshold, the relevant configuration is a localized D0-brane black hole. 
The branch structure is governed by $y = r_L^2 / r_\beta$. 
Numerical construction of these solutions~\cite{Dias:2017uyv} shows a first-order transition at $r_L^2 = \cgrav \, r_\beta$, with $\cgrav \simeq 2.45$.
The homogeneous phase is favored for $r_L^2 > \cgrav r_\beta$ and the localized phase for $r_L^2 < \cgrav r_\beta$. 

For $y \ll 1 < \cgrav$ the localized \emph{D0 phase} admits an analytic expansion of the free energy density~\cite{Harmark:2004ws},
\begin{equation}
\label{eq:D0phase}
\frac{f_{\text{loc}}}{N^2 \la} = - \kappa \frac{t^{14/5}}{r_L^{2/5}} \left(1 + \mathcal{O}(y^{14/5})\right),
\end{equation}
where $\kappa$ is a known constant. 
In the region where this phase dominates, the leading term in \eq{eq:D0phase} accurately captures the thermodynamics. 

All these solutions have contractible thermal circles, implying $P_\beta \neq 0$. 
The spatial behavior distinguishes the phases: in the D1 phase, the spatial circle is non-contractible, giving $P_L = 0$ (spatial confinement), while in the D0 phase the horizon is localized and the eigenvalue distribution of $\cP e^{i\oint_L A}$ is compactly supported~\cite{Kol:2002xz, Aharony:2004ig}, yielding $P_L \neq 0$. 
Thus, the gravity transition at $r_L^2 = \cgrav r_\beta$ corresponds to a first-order spatial deconfinement transition in the gauge theory.

We focus on $r_\beta, r_L \sim \mathcal{O}(1)$ in the large-$N$ limit. 
For parametrically larger $r_\beta$, the supergravity approximation breaks down due to strong curvature near the horizon.

\begin{figure}[tbp]
\centering
\includegraphics[width=0.4\linewidth]{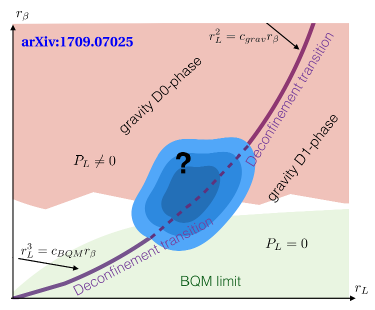}
\caption{\label{fig:summaryrect}Summary of the expected phase structure for 2D MSYM on a rectangular torus. This figure is adapted from Ref.~\cite{Catterall:2017lub}.}
\end{figure}

\subsection{Summary for SYM on a rectangular torus}

We now summarize the expected phase structure of large-$N$ 2D MSYM formulated on a \emph{rectangular} Euclidean torus. 
The system is characterized by the two dimensionless parameters $r_\beta$ and $r_L$, which are held fixed in the large-$N$ limit, $r_\beta, r_L \sim \cO(1)$. 

Combining perturbative arguments, dimensional reduction limits, and holographic predictions, we expect the following regimes.

\textbf{High-temperature, small-volume regime} ($r_\beta, r_L \ll 1$): In this limit, the dynamics are dominated by scalar and gauge zero modes. 
The theory effectively reduces to a matrix integral, leading to $P_\beta \neq 0$ and $P_L \neq 0$, with the bosonic action expectation value approaching $\langle S_{\text{Bos}} \rangle \simeq -2 N^2$. 
Both thermal and spatial center symmetries are therefore broken.

\textbf{High-temperature, large-spatial-circle regime} ($r_\beta^3 \ll r_L$): Here, the theory reduces to bosonic quantum mechanics (BQM) along the spatial direction. 
Thermal deconfinement persists, $P_\beta \neq 0$, while the spatial behavior depends on the relative scaling of $r_L$ and $r_\beta$. 
A spatial deconfinement transition is expected near $r_L^3 = c_{\text{BQM}} \, r_\beta$ where $c_{\text{BQM}} \simeq 1.4$.
For $r_L^3 < c_{\text{BQM}} r_\beta$ one expects $P_L \neq 0$, whereas for $r_L^3 > c_{\text{BQM}} r_\beta$ the spatial center symmetry is restored and $P_L = 0$.

\textbf{Low-temperature gravity regime} ($r_\beta \gg 1$): In this regime, the thermodynamics is governed by the dual supergravity description. 
The free energy depends on the combination $y = r_L^2 / r_\beta$. 
Throughout this regime, the thermal circle is contractible in the bulk geometry, implying $P_\beta \neq 0$. 
As discussed above, a first-order spatial deconfinement transition occurs at $r_L^2 = \cgrav \, r_\beta$ with $\cgrav \simeq 2.45$~\cite{Dias:2017uyv}. 
For $r_L^2 > \cgrav r_\beta$, the homogeneous D1 phase dominates, with free energy density given by Eq.~\eqref{eq:D1phase} and spatial confinement, $P_L = 0$. 
For $r_L^2 < \cgrav r_\beta$, the localized D0 phase dominates, with spatial deconfinement, $P_L \neq 0$, and free energy density given by Eq.~\eqref{eq:D0phase}. 

The resulting phase structure in the $(r_\beta,r_L)$ plane is illustrated schematically in Fig.~\ref{fig:summaryrect}. 

\subsection{Behaviour on a skewed torus} 
\label{sec:skewed} 

We now consider 2D MSYM formulated on a \emph{skewed} Euclidean two-torus. 
This generalization is motivated by two considerations. 
First, varying the skewing parameter provides an additional continuous deformation of the theory and therefore offers an independent test of holography in a regime where gravity predictions remain available. 
Second, modern lattice constructions of supersymmetric gauge theories naturally employ non-hypercubic $A_d^*$ lattices, which correspond in the continuum to field theories defined on skewed tori.

We parametrize the skewing by the dimensionless parameter $\ga = \vec \beta \cdot \vec L / (\beta L)$, with $-1 < \ga < 1$ where $\ga = 0$ is the rectangular case.
While the $A_2^*$ lattice discussed below corresponds to $\ga = -1 / 2$, for certain aspect ratios $\al = r_L / r_\beta$, a modular transformation is required to return to the fundamental domain.
This changes $\ga \to \ga'$, with some aspect ratios producing $\ga' = 0$ while others give $\ga' \ne 0$~\cite{Catterall:2017lub}.
At large $N$, and for fixed $\gamma$, we again expect a phase diagram qualitatively similar to that of the rectangular torus, with a spatial deconfinement transition characterized by the order parameter $P_L$.

\textbf{High-temperature, small-volume regime.} For $r_\beta, r_L \ll 1$, the dynamics are dominated by scalar and gauge zero modes. 
As in the rectangular case, we expect spatial deconfinement, $P_L \neq 0$, and the bosonic action density depends on the aspect ratio~\cite{Catterall:2017lub}:
\begin{equation}
\frac{\langle s_{\text{Bos}}\rangle}{N^2 \lambda} = - \frac{2}{\al \sqrt{1-\ga^2}} \, t^2.
\end{equation}

\textbf{High-temperature dimensional reduction regime.} When $r_\beta^3 \ll r_L$, the theory again reduces to BQM.
In this regime, we expect a spatial deconfinement transition, with $c_{\text{BQM}}$ also modified by $\ga$, so that $P_L \neq 0$ for
\begin{equation}
r_L^3 \lesssim \frac{1.4}{(1 - \gamma^2)^{3/2}}\, r_\beta.
\end{equation}

\textbf{Low-temperature gravity regime.} For sufficiently low temperatures, $t \ll 1$ (equivalently $r_\beta \gg 1$), the thermodynamics is governed by the dual supergravity description. 
As in the rectangular case, two competing phases arise: a spatially localized D0 phase and a homogeneous D1 phase.

The spatially deconfined D0 phase dominates when $r_L^2 < \cgrav \sqrt{1 - \gamma^2} \, r_\beta$.
Its bosonic action density admits the approximate behavior
\begin{equation}
\frac{s_{\text{Bos,D0}}}{N^2 \lambda} = - C_0 \frac{t^{16/5}}{\alpha^{2/5}(1 - \gamma^2)^{7/5}} \left[1 - C_1 \left(\frac{\alpha^2}{t}\right)^{14/5} + \mathcal{O} \, \left( \left( \frac{\alpha^2}{t} \right)^{28/5} \right) \right],
\end{equation}
where $C_0$ and $C_1$ are known numerical constants~\cite{Catterall:2017lub}.

For $r_L^2 > \cgrav \sqrt{1-\gamma^2} \, r_\beta$, the homogeneous D1 phase dominates. 
In this phase, the spatial center symmetry is restored, $P_L = 0$. 
The bosonic action density behaves as
\begin{equation}
\frac{s_{\text{Bos,D1}}}{N^2 \lambda} = - \frac{2^3 \pi^{5/2}}{3^4} \, t^3 \simeq - 1.728 \, t^3,
\end{equation}
where this expression assumes $(r_\beta, r_L, \gamma)$ lie within the fundamental domain.

\textbf{Thermal behavior and phase trajectories.} In all of the above regimes, the gravity solutions possess contractible thermal circles.
Hence, we expect $P_\beta \neq 0$ throughout the parameter space, again assuming $(r_\beta, r_L, \gamma)$ lie in the fundamental domain.

\begin{figure}[tbp]
\centering
\includegraphics[height=4 cm]{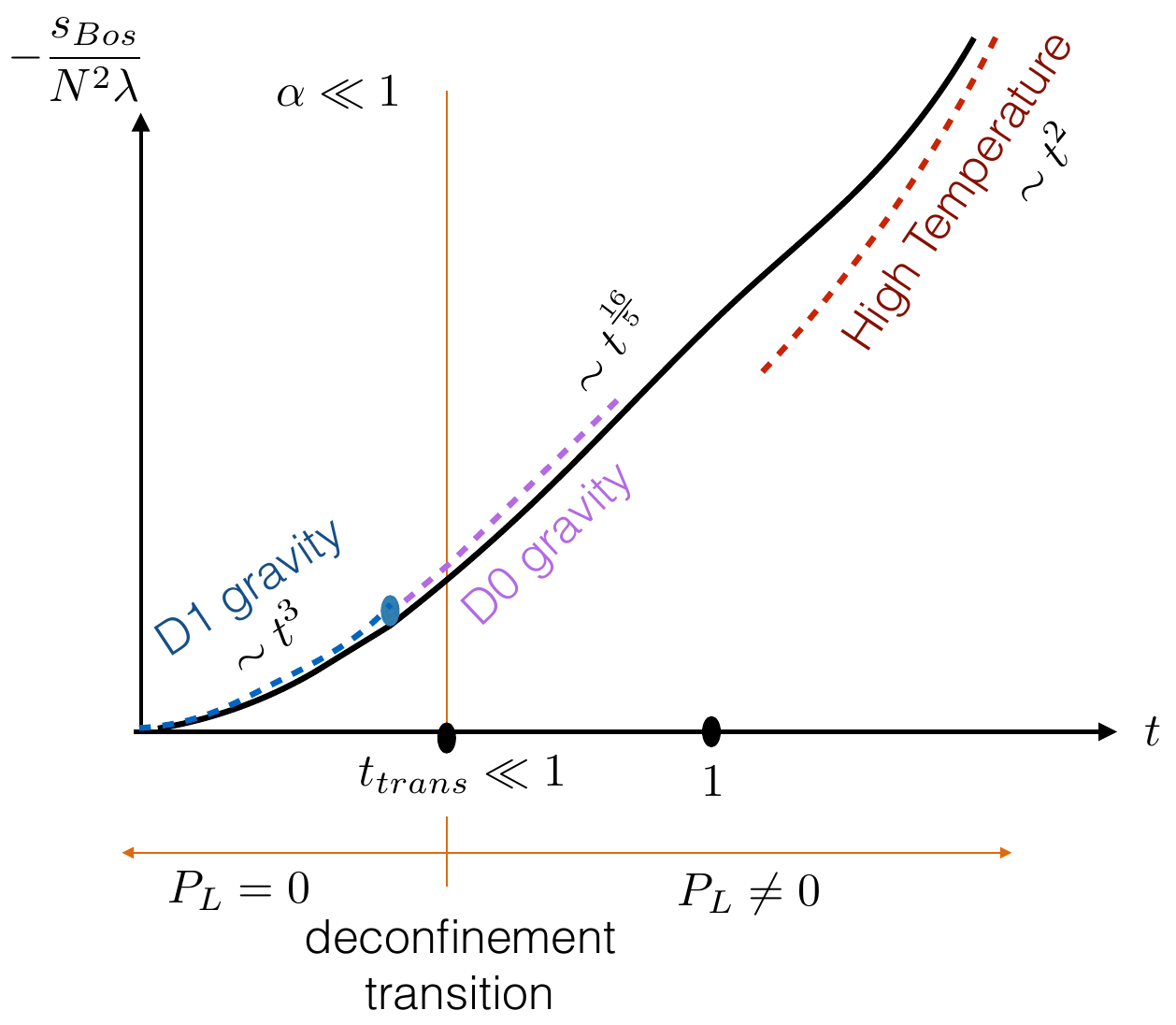}\hfill \includegraphics[height=4 cm]{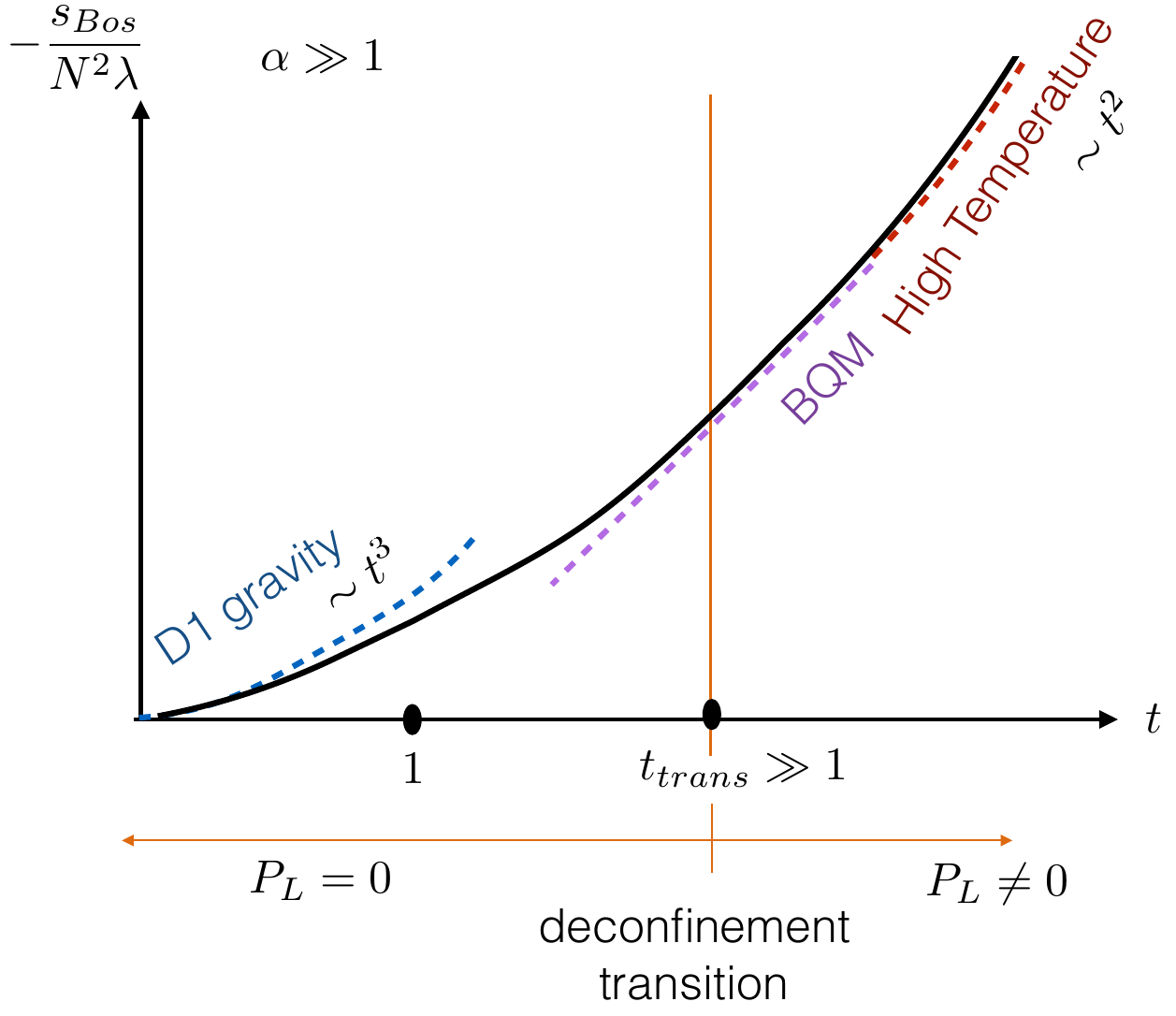}\hfill \includegraphics[height=4 cm]{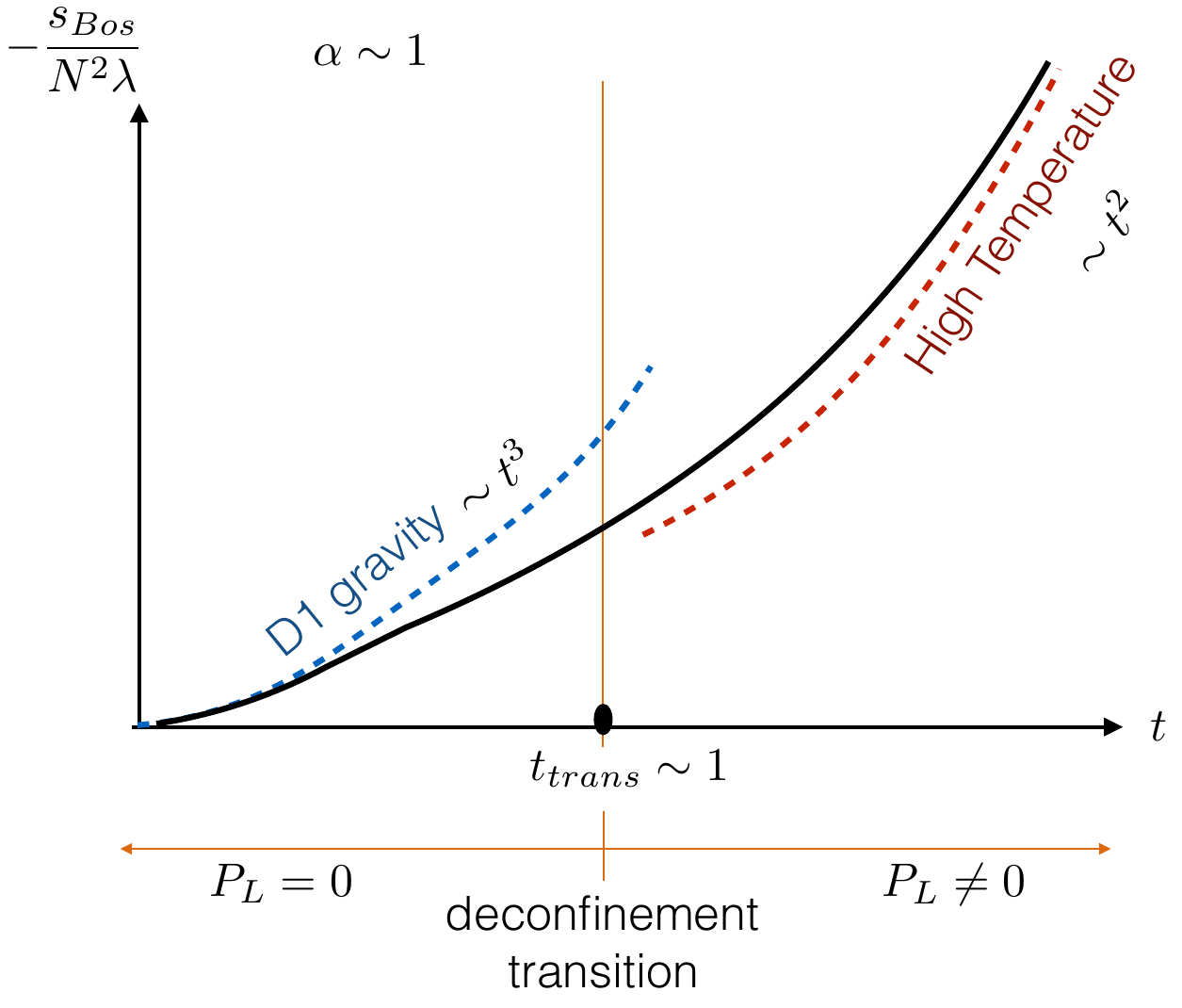}
\caption{\label{fig:summaryskew}Summary of the expected behavior of 2D MSYM on a skewed torus varying $t$ with fixed shape parameters \al and $\ga$. The three plots consider $\al \ll 1$ (left), $\al \gg 1$ (center) and $\al \sim 1$ (right).  These figures are taken from Ref.~\cite{Catterall:2017lub}.}
\end{figure}

For numerical investigations, it is convenient to fix \al and vary the temperature $t$, thereby scanning a one-parameter slice of the $(r_\beta, r_L)$ plane. 
For any finite $\alpha$, sufficiently high temperature ($t \gg 1$) places the system in the small-volume regime with $P_L \neq 0$. 
Upon decreasing $t$, one generically encounters a spatial confinement transition. 
The precise nature of this transition depends on $\alpha$: (i.) For $\alpha \gg 1$, the transition is governed by the BQM regime; (ii.) For $\alpha \ll 1$, the system first enters the spatially deconfined D0 phase, and then undergoes the first-order dual GL transition to the D1 phase at $t = \al^2 / (\cgrav \sqrt{1-\ga^2})$; and (iii.) Intermediate values of \al interpolate smoothly between these behaviors. 
These expectations for large, small, and intermediate \al are illustrated schematically in Fig.~\ref{fig:summaryskew}.

\section{2D MSYM from dimensional reduction}

We can construct supersymmetric lattice gauge theories using ideas from either topological field theory or orbifold projections.
These developments make it possible to discretize $d$-dimensional SYM theories with at least $2^d$ supersymmetries while exactly preserving a subset of the supersymmetries at non-zero lattice spacing. 
The full supersymmetric theory is then recovered in the continuum limit.
See Ref.~\cite{Schaich:2022xgy} for a recent review.

Maximally supersymmetric Yang--Mills theory in any dimension may be obtained by classical dimensional reduction of ten-dimensional $\cN = 1$ SYM. 
In particular, 4D $\cN = 4$ SYM appears on the way from ten dimensions to 2D MSYM. 
We therefore begin with the twisted 4D theory and dimensionally reduce it to obtain the 2D theory of interest.
(Refs.~\cite{Giguere:2015cga, Kadoh:2017mcj} take a different approach also based on topological twisting.)
In this section, we will discuss this procedure in the continuum, followed by lattice discretization in the next section.

A notable feature of twisted 4D $\cN = 4$ SYM is that the gauge field and six adjoint scalar fields are organized into a five-component complexified gauge field $\cA_b$ with $b = 1, \cdots, 5$.
After dimensional reduction, the components of $\cA_b$ along the reduced directions become complexified scalar fields in two dimensions. That is, $\cA_4 \equiv \varphi$ and $\cA_5 \equiv \phi$. The remaining components $\cA_i$ with $i = 1, 2, 3$ contain both the 2D gauge field and the remaining four scalar fields.
The complexified field strength decomposes in the following way: $\cF_{ab} \; \longrightarrow \; \big\{ \cF_{ij},~ \cD_i \phi,~ \cD_i \varphi,~ [\varphi, \phi] \big\}$.

The bosonic part of the 2D continuum action becomes
\begin{equation}
\begin{aligned}
S^{D=2}_B = \frac{N}{2 \lambda} \int d^2x \, {\rm Tr} \, \Big(
& - \overline{\cF}_{ij} \cF_{ij} - 2 \, \overline{\cD_i \phi} \, \cD_i \phi - 2 \, \overline{\cD_i \varphi} \, \cD_i \varphi - 2 \, \overline{[\varphi, \phi]} \, [\varphi, \phi] \\
& + \frac{1}{2} [\overline{\cD}_i, \cD_i]^2 + \frac{1}{2} [\overline{\phi}, \phi]^2 + \frac{1}{2} [\overline{\varphi}, \varphi]^2 \\
& + [\overline{\phi}, \phi] [\overline{\varphi}, \varphi] + [\overline{\cD}_i, \cD_i] [\overline{\phi}, \phi] + [\overline{\cD}_i, \cD_i] [\overline{\varphi}, \varphi] \Big),
\end{aligned}
\end{equation}
where \la is the dimensionful 2D ’t~Hooft coupling.

The twisted fermions are similarly decomposed under dimensional reduction. 
The five-component vector and antisymmetric tensor fermions reorganize in the following way: $\psi_a \; \rightarrow \; \{\psi_i,~ \overline{\zeta},~ \overline{\eta} \}$ and $\chi_{ab} \; \rightarrow \; \{\chi_{ij},~ \overline{\psi}_i,~ \overline{\theta}_i,~ \zeta \}$. All fields transform in the adjoint representation.
The resulting fermionic part of the action is
\begin{equation}
\begin{aligned}
S^{D=2}_F = - \frac{N}{2 \lambda} \int d^2x \, {\rm Tr} \, \Big( 
& \chi_{ij}\,\cD_{[i}\psi_{j]} + 2 \, \overline{\psi}_i \big( \cD_i \overline{\zeta} - [\varphi, \psi_i] \big) + 2 \, \overline{\theta}_i \big(\cD_i \overline{\eta} - [\phi, \psi_i] \big) \\
& + 2 \, \zeta \big([\varphi, \overline{\eta}] - [\phi, \overline{\zeta}] \big) + \eta \, \overline{\cD}_i \psi_i + \eta [\overline{\varphi}, \overline{\zeta}] + \eta [\overline{\phi}, \overline{\eta}] \Big).
\end{aligned}
\end{equation}

In addition to the $\cQ$-exact contributions above, the twisted continuum action of 2D MSYM contains a $\cQ$-closed term inherited from the 4D theory.
After dimensional reduction, this becomes
\begin{equation}
\begin{aligned}
S^{D=2}_{\cQ\text{-closed}} =  \frac{N}{4\lambda} \int d^2x \, {\rm Tr} \, \Big( 
& \varepsilon_{ijk} \, \zeta \, \overline{\cD}_i \chi_{jk} + 2 \, \varepsilon_{ijk} \, \overline{\psi}_i \, \overline{\cD}_j \overline{\theta}_k \\
& + \varepsilon_{ijk} \, \overline{\psi}_i [\overline{\phi}, \chi_{jk}] - \varepsilon_{ijk} \, \overline{\theta}_i [\overline{\varphi}, \chi_{jk}] \Big).
\end{aligned}
\end{equation}

Together, $S^{D=2}_B$, $S^{D=2}_F$ and $S^{D=2}_{\cQ\text{-closed}}$ yield the full continuum action of 2D MSYM in its twisted formulation. 
This continuum theory, defined on a skewed torus with $\ga = - 1/2$, provides the target for the lattice discretization we now discuss.

\section{Lattice discretization}

In both 4D and 2D, the lattice theory is a direct discretization of the twisted continuum theory.
As a result of the five-component formulation discussed above, the 4D theory employs the $A_4^*$ lattice rather than the hypercubic lattice. 
Upon dimensional reduction, this yields a 2D lattice with $A_2^*$ (triangular) geometry. 
Imposing thermal boundary conditions (all periodic except for anti-periodic fermion boundary conditions along one cycle) produces a continuum limit corresponding to 2D MSYM formulated on a skewed torus. 
The skewing arises because the $A_2^*$ basis vectors are not orthogonal. 
For the $A_2^*$ lattice, the skewing parameter is $\gamma = - 1/2$, but this may be changed by a modular transformation as discussed in Sec.~\ref{sec:skewed}.

We follow the geometric discretization prescription given in Ref.~\cite{Damgaard:2008pa} to formulate both theories on the lattice. 
In this approach, continuum complexified covariant derivatives are mapped to finite-difference operators involving complexified link variables, while $p$-form fields are placed on the corresponding $p$-cells of the lattice. 
Curl- and divergence-like operators become gauge-covariant forward and backward difference operators, denoted $\cD_i^{(+)}$ and $\cD_i^{(-)}$, respectively.

After dimensional reduction of twisted 4D $\cN = 4$ SYM, the complexified gauge fields become link variables $\cU_i(\vn)$ living on the links of the 2D $A_2^*$ lattice while the complexified scalar fields $\phi(\vn)$ and $\varphi(\vn)$ reside on lattice sites.
The fermions are distributed over sites, links, and plaquettes according to their form degree in the twisted construction. 
This placement ensures exact gauge invariance and preservation of four nilpotent twisted-scalar supersymmetries \cQ at non-zero lattice spacing.

The bosonic part of the lattice action takes the form
\begin{align}
S_B^{\mathrm{lat}} = \frac{N}{2 \lambda_{\text{lat}}} \sum_{\mathbf{n}} {\rm Tr} \, \Big( & - \overline{\cF}_{ij}(\mathbf{n}) \, \cF_{ij}(\mathbf{n}) - 2 \, \overline{\cD_i^{(+)} \varphi(\mathbf{n})} \, \cD_i^{(+)} \varphi(\mathbf{n}) - 2 \, \overline{\cD_i^{(+)} \phi(\mathbf{n})} \, \cD_i^{(+)} \phi(\mathbf{n}) \nonumber \\
& - 2 \, \overline{[\varphi(\mathbf{n}), \phi(\mathbf{n})]} \, [\varphi(\mathbf{n}), \phi(\mathbf{n})] + \frac{1}{2} [\overline{\phi}(\mathbf{n}), \phi(\mathbf{n})]^2  \nonumber \\
& + \frac{1}{2} [\overline{\varphi}(\mathbf{n}), \varphi(\mathbf{n})]^2 + [\overline{\phi}(\mathbf{n}), \phi(\mathbf{n})] [\overline{\varphi}(\mathbf{n}), \varphi(\mathbf{n})] \\
& + \frac{1}{2} \big( \overline{\cD}_i^{(-)} \cU_i(\mathbf{n}) \big)^2 + \overline{\cD}_i^{(-)} \cU_i(\mathbf{n}) [\overline{\phi}(\mathbf{n}), \phi(\mathbf{n})] + \overline{\cD}_i^{(-)} \cU_i(\mathbf{n}) [\overline{\varphi}(\mathbf{n}), \varphi(\mathbf{n})] \Big), \nonumber 
\end{align}
where the lattice field strength $\cF_{ij}$ is constructed from the forward difference operator in the usual gauge-covariant manner.

The fermionic part of the lattice action is
\begin{align}
S_F^{\mathrm{lat}} = - \frac{N}{2 \lambda_{\text{lat}}} \sum_{\mathbf{n}} {\rm Tr} \, \Big( & \chi_{ij}(\mathbf{n}) \, \cD^{(+)}_{[i}\psi_{j]}(\mathbf{n}) + 2 \, \overline{\psi}_i(\mathbf{n}) \Big( \cD^{(+)}_i \overline{\zeta}(\mathbf{n}) - [\varphi(\mathbf{n}), \psi_i(\mathbf{n})] \Big) \nonumber \\
& + 2 \, \overline{\theta}_i(\mathbf{n}) \Big( \cD^{(+)}_i \overline{\eta}(\mathbf{n}) - [\phi(\mathbf{n}), \psi_i(\mathbf{n})] \Big) + 2 \, \zeta(\mathbf{n}) \Big( [\varphi(\mathbf{n}), \overline{\eta}(\mathbf{n})] - [\phi(\mathbf{n}), \overline{\zeta}(\mathbf{n})] \Big) \nonumber \\ 
& + \eta(\mathbf{n}) \, \overline{\cD}_i^{(-)} \psi_i(\mathbf{n}) + \eta(\mathbf{n}) [\overline{\varphi}(\mathbf{n}), \overline{\zeta}(\mathbf{n})] + \eta(\mathbf{n}) [\overline{\phi}(\mathbf{n}), \overline{\eta}(\mathbf{n})] \Big).
\end{align}

In addition to the $\cQ$-exact contributions above, the lattice action includes a discretized $\cQ$-closed term inherited from the 4D construction:
\begin{align}
S_{\cQ\text{-closed}}^{\mathrm{lat}} =  \frac{N}{4 \lambda_{\text{lat}}} \sum_{\mathbf{n}} {\rm Tr} \, \Big( & \varepsilon_{ijk} \, \overline{\psi}_i(\mathbf{n}) [\overline{\phi}(\mathbf{n}), \chi_{jk}(\mathbf{n})] - \varepsilon_{ijk} \, \overline{\theta}_i(\mathbf{n}) [\overline{\varphi}(\mathbf{n}), \chi_{jk}(\mathbf{n})] \nonumber \\
& + \varepsilon_{ijk} \, \zeta(\mathbf{n}) \, \overline{\cD}_i^{(+)} \chi_{jk}(\mathbf{n}) + 2 \, \varepsilon_{ijk} \, \overline{\psi}_i(\mathbf{n}) \, \overline{\cD}_j^{(+)} \overline{\theta}_k(\mathbf{n}) \Big).
\end{align}

The action $S_B^{\mathrm{lat}} + S_B^{\mathrm{lat}} + S_{\cQ\text{-closed}}^{\mathrm{lat}}$ possesses flat directions that must be regulated in numerical computations. 
We introduce a soft supersymmetry-breaking scalar potential,
\begin{equation}
\Delta S = \frac{N}{4 \lambda_{\text{lat}}}\,\mu^2 \sum_{\mathbf{n}} \operatorname{Tr}  
\Bigg[
\sum_i \left( \overline{\mathcal U}_i(\mathbf{n}) \mathcal U_i(\mathbf{n}) - \mathbb{I} \right)^2
+ \left( \overline{\phi}(\mathbf{n}) \phi(\mathbf{n}) - \mathbb{I} \right)^2
+ \Big( \overline{\varphi}(\mathbf{n}) \varphi(\mathbf{n}) - \mathbb{I} \Big)^2
\Bigg],
\end{equation}
which stabilizes the scalar fields while vanishing in the $\mu \to 0$ limit. 

The lattice construction preserves exact gauge invariance and (for $\mu = 0$) four nilpotent twisted-scalar supersymmetries $\cQ$ at finite lattice spacing. 
The full $\cN = (8, 8)$ supersymmetry is expected to be restored in the continuum limit without fine-tuning.

We will carry out our numerical computations using \texttt{SUSY LATTICE}, parallel software for lattice studies of supersymmetric Yang--Mills theories in various dimensions.
The code is based on the MILC software, and its design, parallelization strategy, and numerical algorithms are presented in Ref.~\cite{Schaich:2014pda}.  
We are currently extending the existing code base to incorporate the 2D MSYM lattice formulation detailed above~\cite{susy_code}.

\section{Conclusions and Future Directions}

In this work, we have outlined our program for numerical studies of 2D MYSM formulated on a (rectangular or skewed) Euclidean torus, with anti-periodic (thermal) boundary conditions imposed on fermions along one cycle. 
The theory is characterized by three dimensionless parameters: the spatial and temporal scales $r_L$, $r_\beta = 1/t$, and the skewing parameter \ga specifying the torus geometry.

According to holography, at low temperature $t \ll 1$ and large $N$, this theory provides a non-perturbative description of a dual supergravity.
There are two competing phases in the gravity phase diagram: a homogeneous D1 (black string) phase and a localized D0 (black hole) phase. 
The transition line separating these phases is conjectured to correspond to a first-order spatial deconfinement transition in the gauge theory. 
The spatial Wilson line $P_L$ serves as the order parameter.

We formulate 2D MSYM on the triangular lattice using a twisted construction that preserves exact gauge invariance and four nilpotent twisted-scalar supersymmetries at finite lattice spacing. 
The supersymmetric discretization fixes the skewing parameter to $\gamma = -1/2$ up to modular transformations that may be required to return to the fundamental domain~\cite{Catterall:2017lub}.
This allows us to study both rectangular and skewed geometries, thereby providing additional tests of holography.

Our objective is to map out the non-perturbative phase diagram as a function of temperature and aspect ratio, determine the transition line between spatially confined and deconfined phases, and compare its parametric form with supergravity predictions. 
We will also compute thermodynamic observables such as the bosonic action density and test their consistency with the corresponding black hole thermodynamics at low temperature. 
This framework enables a systematic numerical exploration of holography on both rectangular and skewed tori, which will surpass previous studies and offer new tests of gauge/gravity duality.

\acknowledgments

The work of A.J.\ was supported in part by a Start-up Research Grant from the University of the Witwatersrand.
B.S.S.\ gratefully acknowledges the Institute Fellowship provided by the Indian Institute of Science Education and Research (IISER) Mohali for financial support during the Ph.D.\ program. 
D.S.\ is supported by UK Research and Innovation Future Leader Fellowship {MR/X015157/1} as well as Science and Technology Facilities Council consolidated grant {ST/X000699/1}.
Computational resources were provided by the Param Smriti High-Performance Computing facility, supported by the Government of India. \\[8 pt] 

\noindent \textbf{Data Availability Statement:} No new data have been generated for this proceedings.  \refcite{susy_code} provides the code where we are implementing the presented lattice formulation of 2D MSYM.

\bibliographystyle{JHEP}
\bibliography{lattice25}
\end{document}